# SNS SUPERCONDUCTING CAVITY MODELING -ITERATIVE LEARNING CONTROL

Sung-il Kwon, Yi-Ming Wang, Amy Regan, Tony Rohlev,
LANL, Los Alamos, NM87544, USA
Mark Prokop, Dave Thomson, Honeywell FM&T

**Abstract**

The SNS SRF system is operated with a pulsed beam. For the SRF system to track the repetitive reference trajectory, a feedback and a feedforward controllers has been proposed. The feedback controller is to guarantee the closed loop system stability and the feedforward controller is to improve the tracking performance for the repetitive reference trajectory and to suppress the repetitive disturbance. As the iteration number increases, the error decreases.

## 1 INTRODUCTION

The Spallation Neutron Source (SNS) Linac to be built at Oak Ridge National Laboratory (ORNL) consists of a combination of low energy normal conducting (NC) accelerating structures as well as higher energy superconducting RF (SRF) structures. In order to efficiently provide a working control system, a lot of modeling has performed. The modeling is used as a way to specify RF components; verify system design and performance objectives; optimize control parameters; and to provide further insight into the RF control system operation.

The modeling addressed in this note deals with the PI feedback controller and the plug-in feedforward controller (the iterative learning controller). The purpose of the PI feedback controller is to guarantee the robustness and the zero steady state error. However, the PI feedback controller does not yield the satisfactory transient performances for the RF filling and the beam loading. The feedforward controller proposed in this note takes a simple form and is effective. In order to generate the one step ahead feedforward control, the feedforward controller makes use of current error, the derivative of the current error and the integration of the current error. This PID-type feedforward controller is the natural consequence of the PI feedback control system where the inverse of the closed loop system transfer matrix has the same form as the transfer matrix of the PID system. The proposed feedforward controller achieves the better performance for the repetitive reference trajectory to be tracked by the system output and achieves the suppression of the repetitive disturbance such as the Lorentz force detuning.

## 2 SUPERCONDUCTING CAVITY MODEL

The modeling of a superconducting cavity is based on the assumption that the RF generator and the cavity are connected with a transformer. The equivalent circuit of the cavity is transformed to the equivalent circuit of RF generator with transmission line (wave guide) and the model is obtained[2]. A superconducting cavity is represented by the state space equation.

$$\dot{x} = A(\Delta\omega_L)x + B(\Delta\omega_L)u + B_I(\Delta\omega_L)I \quad (1)$$
$$y = C(\Delta\omega_L)x$$

and the Lorentz force detuning is

$$\dot{\Delta\omega}_L = -\frac{1}{\tau_m}\Delta\omega_L - \frac{2\pi}{\tau_m}\bar{K}z_1^2 - \frac{2\pi}{\tau_m}\bar{K}z_2^2 \quad (2)$$

where

$$A(\Delta\omega_L) = \begin{bmatrix} -\frac{1}{\tau_L} & -(\Delta\omega_m + \Delta\omega_L) \\ (\Delta\omega_m + \Delta\omega_L) & -\frac{1}{\tau_L} \end{bmatrix}, \quad C = \begin{bmatrix} 1 & 0 \\ 0 & 1 \end{bmatrix},$$

$$B(\Delta\omega_L) = \begin{bmatrix} \frac{2}{Z_o}c_1 & -\frac{2}{Z_o}c_3 \\ \frac{2}{Z_o}c_3 & \frac{2}{Z_o}c_1 \end{bmatrix}, \quad B_I(\Delta\omega_L) = \begin{bmatrix} -2c_1\zeta & 2c_3\zeta \\ -2c_3\zeta & -2c_1\zeta \end{bmatrix},$$

$$c_1 = \frac{R_{cu}}{\tau}, \quad c_3 = \frac{R_{cu}}{2Q_o\tau}, \quad \bar{K} = K\left[\zeta\frac{E_o[MV/m]}{V_{gap}[V]}\right]^2$$

$\zeta$ : Transformation ratio, $\quad Q_o$ : Unloaded $Q$

$R_{cu}$ : Resistance of the cavity equivalent circuit

$\Delta\omega_m$ : Detuning frequency[rad/s]

$Z_o$ : Transmission line impedance

$\tau_L$ : Loaded cavity damping constant
$\tau$ : Unloaded cavity damping constant
$\tau_m$ : Mechanical time constant
$K$ : Lorentz force detuning Constant
$u = \begin{bmatrix} V_{fI} & V_{fQ} \end{bmatrix}^T$ : forward Voltage in I/Q
$I = \begin{bmatrix} I_I & I_Q \end{bmatrix}^T$ : Beam current in I/Q
$x = \begin{bmatrix} V_I & V_Q \end{bmatrix}$ : Cavity Field in I/Q

The modeling of the cavity is based on the assumption that the exact characteristics, parameters of a cavity are known. When there are parameter perturbations, unknown deterministic disturbances and random noises in the input channels or measurement channels, those uncertainties are added to the state equation or the output equation. For the control of this uncertain system, modern robust controllers such as $H_\infty$ controller, loop-shaping controller are applied. On the other hand, PI (PID) controllers are designed by using $H_\infty$ controller, loop-shaping controller design techniques.

## 3 ITERATIVE LEARNING CONTROL

The SNS SRF system is operated with a pulsed beam. The period of the beam pulse is 16.67 $m\sec$ ($1/60$ $Hz$). The objective of the SRF controller is to generate a periodic reference trajectory whose period is 16.67 $m\sec$ ($1/60$ $Hz$) and is to achieve a stable cavity field periodically so that the RF power is delivered to the periodic beam pulse safely[3]. A control system that is suited for this type of applications is Iterative Learning Control (ILC) [1],[3].

Consider a controller at the *kth* iteration,

$$u^k = u_C^k + u_F^k \quad (3)$$

where $u_C^k$ is the output of the PI feedback controller and $u_F^k$ is the output of the feedforward ILC controller. The error dynamics is expressed as

$$\dot{e}^k = A_c e^k - BK_I x_c^k$$
$$\quad - Bu_F^k - B_I I_B - A(\Delta \omega_L)r + \dot{r} \quad (4)$$
$$\dot{x}_c^k = e^k$$

where $A_c = A(\Delta \omega_L) - BK_P$. Since $c_1 \gg c_3$, with the proper diagonal terms and zero off-diagonal terms of the gain matrices $K_P$ and $K_I$ of the PI controller, the diagonal terms of the matrix $A(\Delta \omega_L) - BK_P$ and the matrix $BK_I$ are sufficiently large and so the I channel error and the Q channel error (4) are almost decoupled.

The Laplace transform of the error equation (4) yields

$$E^k(s) = -S_e(s)U_F^k(s) - S_e(s)B^{-1}B_I I_B(s)$$
$$\quad + S_e(s)B^{-1}(sI - A(\Delta \omega_L))R(s) \quad (5)$$

where

$$S_e(s) = \left(sI - A_c + \frac{1}{s}BK_I\right)^{-1} B \quad (6)$$

Define the learning control rule as follows.

$$U_F^{k+1} = Q\left(f \cdot U_F^k + \alpha \cdot LE^k\right) \quad (7)$$

where $f$, $0 < f < 1$, is called the forgetting factor and $\alpha$, $0 < \alpha < 1$, is a design constant. The forgetting factor $f$ and the constant $\alpha$ are to guarantee the robust stability against uncertainties in the plant model and the nonlinearity of the klystron. They also allow for elimination of the influence of random noise, spikes and glitches. $U_F^k$ is the Laplace transform of the feedforward signal in iteration $k$ and $E^k$ is the Laplace transform of the corresponding tracking error. Learning converges if the feedback loop is stable and the following condition holds. For $\forall \omega \in \Re$,

$$\left\| U_F^{k+2}(j\omega) - U_F^{k+1}(j\omega) \right\|_\infty < \left\| U_F^{k+1}(j\omega) - U_F^k(j\omega) \right\|_\infty,$$

which results in learning convergence condition

$$\left\| Q(f \cdot I - \alpha \cdot LS_e) \right\|_\infty < 1 \quad (8)$$

The $Q$-filter is designed such that it suppresses the high frequency components at which the plant model is inaccurate and passes low frequency, at which the model is accurate. The $Q$-filter is either placed before the memory, or in the memory feedback loop. Thus, the bandwidth of the $Q$-filter should be chosen greater than or equal to the desired closed loop bandwidth. From the $H_\infty$ controller design point of view, (8) interprets the - $Q$-filter as a weighting function for learning performance, i.e.,

$$\|f \cdot I - \alpha \cdot LS_e\|_\infty < \|Q^{-1}\|_\infty \qquad (9)$$

It seems natural that the $Q$-filter is viewed as a measure of learning performance and the cut-off frequency $\omega_c$ of the $Q$-filter is chosen as large as possible in order to guarantee zero tracking error up to frequency $\omega_c$.

To design a $L$-filter, detailed knowledge of the plant is required. For low frequency dynamics, a competent model of the plant often exists. However, identification and modeling of high frequency dynamics is difficult and may lead to an inadequate model. This could result in a learning filter $L$ that compensates well for low frequencies but does not compensate appropriately for all high frequencies and therefore causes unstable behavior. This unstable behavior is prevented by the $Q$-filter and to determine $\omega_c$, a trade-off between the performance and the robust stability is necessary. An intuitive synthesis of the learning $L$-filter for given $Q$-filter is as follow.

$$L(s) = S_e^{-1}(s) = \left(sI - A_c + \frac{1}{s}BK_I\right)B^{-1} \qquad (10)$$

When the feedback PI controller gain matrix $K_I$ is defined as a diagonal matrix, then (10) is reduced to

$$L(s) = sB^{-1} - (A(\Delta\omega_L)B^{-1} - K_P) + \frac{1}{s}K_I \qquad (11)$$

Equation (11) shows that the learning L-filter has the characteristics of PID[3].

## 4  SIMULATION

The closed loop system with PI feedback controller and iterative learning controller was simulated. Figure 1 and figure 2 show the field amplitude and the field phase, where the great improvement of the transient behaviors both in RF filling and in beam loading is observed as iteration number increases. Also, two figures show that the periodic Lorentz force detuning effect on the field amplitude and the field phase is suppressed gradually as the iteration number increases. Figure 3 shows the Lorentz force detuning. Note that the static value of the Lorentz force detuning calculated with the cavity data ( $K = -2.0$ Hz/(MV/m)$^2$, $E_{acc} = 11.9$ MV/m ) is -283 $Hz$. With the RF On period 1.3 msec (300 $\mu$ sec field settling period + 1000 $\mu$ sec beam period), the Lorentz force detuning is developed up to –200 $Hz$.

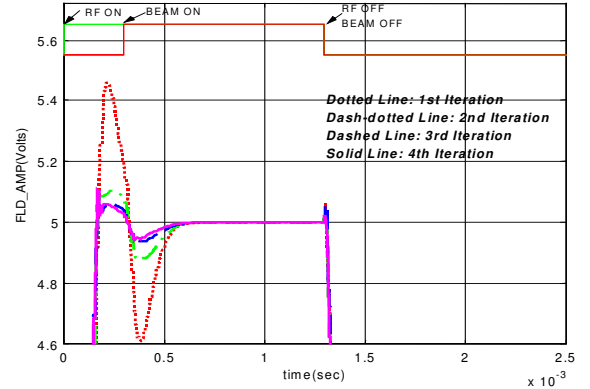

**Figure 1** Field Amplitude with PI Controller plus Iterative Learning Controller (PI+ILC).

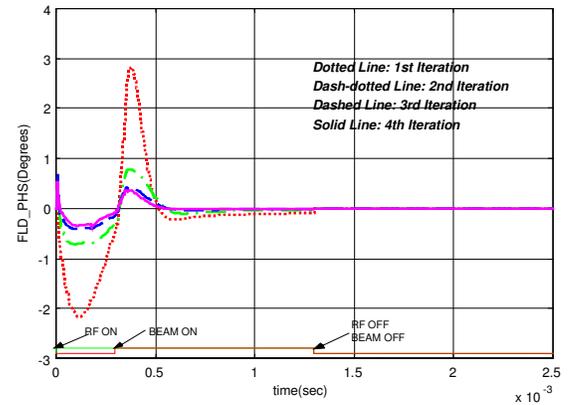

**Figure 2** Field Phase with PI Controller plus Iterative Learning Controller (PI+ILC).

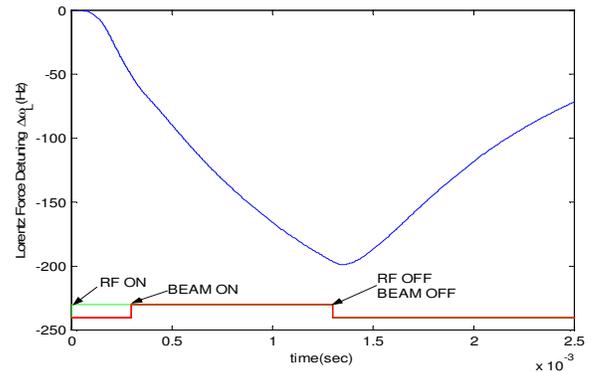

**Figure 3** Lorentz Force Detuning.